\documentclass[hvmath]{copernicus2}

\usepackage{url,lineno}
\usepackage{color}
\usepackage[T1]{fontenc}

\begin{document}

\title{Plasma wave mediated electron pairing effects%
}

\author[1,2]{R. A. Treumann\thanks{Visiting the International Space Science Institute, Bern, Switzerland}}
\author[3]{W. Baumjohann}

\affil[1]{Department of Geophysics and Environmental Sciences, Munich University, Munich, Germany}
\affil[2]{Department of Physics and Astronomy, Dartmouth College, Hanover NH 03755, USA}
\affil[3]{Space Research Institute, Austrian Academy of Sciences, Graz, Austria}

\runningtitle{Electron Pairing}

\runningauthor{R. A. Treumann and W. Baumjohann}

\correspondence{R. A.Treumann\\ (rudolf.treumann@geophysik.uni-muenchen.de)}

\received{ }
\revised{ }
\accepted{ }
\published{ }


\firstpage{1}

\maketitle

\section*{Abstract}
Pairing of particles, in particular electrons, in high temperature plasma is generally not expected to occur. Here we investigate, based on earlier work, the possibility for electron pairing mediated in the presence of various kinds of plasma waves. We confirm the possibility for pairing in ion- and electron-acoustic waves, pointing out the importance of the former and the expected consequences. While electron-acoustic waves probably do not play any role, ion-acoustic waves may cause formation of heavy electron compounds. Lower hybrid waves also mediate pairing but under different conditions. Buneman modes which evolve from strong currents may cause pairing among trapped electrons constituting a heavy electron component that populates electron holes. All pairing processes are found to generate cold pair populations. They provide a mechanism of electron cooling which can be interpreted as kind of classical condensation, in some cases possibly accompanied by formation of current filaments, weak soft-X-ray emission and superfluidity which might affect reconnection physics. 

\introduction
Plasmas consist of equal numbers of electrons and ions forming quasi-neutral fluid-like matter at temperatures sufficiently high for maintaining ionization. At such high temperatures electrons and ions are mutually well separated located at instantaneous distances $L_N\sim N^{-1/3}$, where $N\equiv N_e=N_i$ is the average plasma density in a singly charged plasma. The electrostatic Coulomb fields of the naked electric charges are confined to Debye spheres of radius $\lambda_D$ by the collective effect of the many particles of opposite charge passing around at their average tangential speeds within radial distances $r\lesssim \lambda_D$. The geometric shape of the Debye spheres is very close to a sphere, deviating from it only in very strong magnetic fields and for very high plasma flow speeds. Outside the Debye sphere the residual particle field decays exponentially while contributing to a thermal fluctuation background field. From a particle point of view each of the plasma particles is a charged Fermion. In classical plasmas at the high plasma temperatures the spin has no importance, and the fermionic property of the particles plays no role. In quantum plasmas which, for obeying quantum properties must be dense, this property is rather important. For, when two electrons form pairs, the spins add up and the pair becomes a Boson of either zero or integer spin. Many pairs can occupy the same energy level and, altogether, tend to condensate in the lowest energy level permitted by the temperature. This property is very well known from solid state physics \citep[cf., e.g.,][]{fetter1971,huang1987}. 

In plasmas pairing is a property which is not expected under normal conditions. However, when a plasma wave passes across the plasma, the dielectric properties of the plasma change. Plasma electrons assuming a relative velocity with respect to the phase velocity of the plasma wave find themselves exposed to the dielectric polarization which adds to the Debye screening that compensates for the naked particle charge. Such electrons evolve an attractive electrostatic interparticle force acting on its neighbor electrons, an effect different from classical wave trapping in the wave potential trough that causes wave saturation and other nonlinear effects like deformation, solitons and holes. The attractive forces are dc forces. Experienced by two electrons of approximately same velocity they bind these together to form pairs. Experienced by many electrons of same velocity they can form large compounds of particles the nature of which is that of massive macro-particles of same charge-to-mass ratio $e/m$. Depending on the number $n$  of particles in the conglomerate being odd or even, macroparticles behave like Fermions or Bosons not only following different statistics but  splitting the plasma into two different populations of different energies. Bosons may, in principle, condensate to form a dense and cool population. If this happens it may have a profound effect on a some plasma processes.        

\section{Generation of wave-mediated attractive potentials}
The method of calculating the potential around a test charge in plasma was explicated sixty years ago \citep{neufeld1955}. Thirty years later it was revived \citep{nambu1985} to include the effect of plasma waves and was used in this form to suggest the coagulation of particles in the presence of dust in plasmas in order to explain the formation of dust structure, which for some while became an industry \citep[cf., e.g.,][and references therein]{shukla1997,shukla2001,nambu2001}. Following \citet{neufeld1955}, the general expression for the electrostatic potential $\Phi(\mathbf{x},t)$ is obtained from Poisson's law -$\nabla^2\epsilon(\mathbf{x},t)\Phi(\mathbf{x},t)= qN(\mathbf{x},t)/4\pi\epsilon_0$. The density of a test particle traversing the plasma with velocicty $\mathbf{v}$ and charge $q=q_t$ is $N_t=(2\pi)^3\delta(\mathbf{r})$. In Fourier space this yields for the potential \citep[cf.,][for a textbook]{neufeld1955,krall1973}
\begin{equation}\label{eq-one}
\Phi(\mathbf{x},t)=-\frac{q_t}{8\pi^2\epsilon_0}\int \mathrm{d}\mathbf{k}\,\mathrm{d}\omega\ \frac{\delta(\omega-\mathbf{k\cdot v})}{k^2\epsilon(\mathbf{k},\omega)}\mathrm{e}^{i\mathbf{k\cdot r}}
\end{equation}
Here $\mathbf{r}=\mathbf{x}-\mathbf{v}t$ is the distance between the location $\mathbf{x}'=\mathbf{v}t$ of the particle and the reference point of measurement of the potential disturbance. $\omega(\mathbf{k})$ is the frequency of a spectrum of plasma wave eigenmodes, presumably present in the plasma as background noise or wave excitations, with wave number $\mathbf{k}$.  $\epsilon(\mathbf{k},\omega)$ is the dielectric plasma response function corresponding to the disturbance caused by the test particle. [One may note that the $\delta$-function in the numerator can be used to replace the wave number component parallel to the particle velocity in the Fourier exponential $\exp(i\mathbf{k\cdot  r})$ reducing it to an $\omega$-integration.] 

The problem as seen from the test charge is spherically symmetrical. Thus it makes sense to formulate it in spherical  coordinates $r,k,\Omega_r,\Omega_k$ both in wavenumber and real space, with conventional notation for the angular volume elements. Chosing an expansion into spherical harmonics \citep[as done in][]{neufeld1955} one has for the exponential factor
\begin{equation}
\mathrm{e}^{i\mathbf{k\cdot r} }= 4\pi\sum_l\sum_m i^l\sqrt{\frac{\pi}{2kr}}J_{l+\frac{1}{2}}(kr) Y_l^{m*}(\Omega_k)Y_l^m(\Omega_r).
\end{equation}
The notation for the spherical harmonics $Y_l^m(\Omega_k),Y_l^m(\Omega_r)$  is again conventional in wavenumber and real space, and the $*$ indicates the conjugate complex version of the azimuthal exponentials $\exp(im\phi_k)$. The $k,r$ dependence is taken care of by the half integer Bessel functions $J_{l+{1/2}}(kr)$. 

The response function $\epsilon[\mathbf{k},\omega(\mathbf{k})]$ is a function of frequency and wave number and is taken in the electrostatic limit. In the above representation it is a scalar function. When electromagnetic contributions or an external magnetic field would have to be taken into account, it becomes a tensor. Only its longitudinal part $\epsilon_L=\mathbf{k\cdot{\mathrm\epsilon}\cdot k}/k^2$ enters the expression for the potential, however.  In addition, one would have to consider a variation of the vector potential caused by the test particle, if transverse waves are included. 

In the absence of the latter, it is well known that the potential of the test particle, in our case an electron of elementary charge $q=-e$,  consists of its Coulomb potential $\Phi_C(r)=-q/4\pi\epsilon_0 r$ , which will be Debye-screened by the plasma particles becoming $\Phi_D(r)=\Phi_C(r)\exp(-r/\lambda_D)$, and a disturbance caused by the reaction of the plasma eigenmodes to the presence of the charge -- the eigenmodes that are either present or are amplified by the moving test charge for which the plasma appears as a dielectric to whose normal modes the charge couples. These eigenmodes contribute via adding each of them  to the vacuum dielectric constant its particular susceptibilities $\chi_s(\omega,\mathbf{k})$, where $s$ is the index identifying the particle species which responds to the eigenmodes. Hence, with $s=e,i$ for electrons and ions, respectively,
\begin{equation}\label{eq-potential}
\epsilon(\mathbf{k},\omega) = 1 + (k\lambda_D)^{-2}+\chi_e(\mathbf{k},\omega) + \chi_i(\mathbf{k},\omega)
\end{equation}
Independent of the wave modes, the Debye term on the right is included here in order to account for the presence of the point like  test charge. In a non-magnetized plasma the susceptibilities assume the form
\begin{eqnarray}
\chi_s(\mathbf{k},\omega) &=& (k\lambda_{Ds})^{-2}\left[1+\zeta_sZ(\zeta_s)\right], \\ \zeta_s &=& (\omega-\mathbf{k\cdot u}_s)/kv_s\nonumber
\end{eqnarray}
$\lambda_{Ds}$ is the Debye length of species $s$, $Z(\zeta_s)$ the plasma dispersion function, $v_s$ the thermal speed of species $s$, and $\mathbf{u}_s$ is a possible bulk streaming velocity of species $s$ which, for our application, will for simplicity be put to zero but has to be retained both for streaming and electric currents $\mathbf{J}=\sum_sq_sN_s\mathbf{u}_s$. One should note that Poisson's law is quite general holding for both linear and nonlinear interactions. Restriction to linear response functions only implies small disturbances caused. For large nonlinear disturbances the linear response function would have to be replaced by its nonlinear counterparts.

 These expressions have been partially analyzed in the available literature with focus on the effect of dust in plasmas \citep[adding the susceptibility of dust particles in][and others]{shukla1997,shukla2001}. In the following we follow some of the lines in these papers in view of application to space plasma conditions and with the intention of checking the chances for electron pairing and possible pairing effects.

\subsection{Ion-acoustic pairing potential
}
Our first example is the response of the test charge potential to the presence of a spectrum of ion acoustic waves in a plasma, a problem originally treated cursorily by \citet{nambu1985}.  In this case the linear response function is well known. Neglecting damping, its real part is given by
\begin{equation}\label{eq-respia}
\epsilon_\mathit{ia}(\omega,\mathbf{k})=1+\frac{1}{k^2\lambda_{De}^2}-\frac{\omega_i^2}{\omega^2}\Big(1+3k^2\lambda_{Di}^2\Big)
\end{equation}
where in the round brackets we iterated the frequency by approximating it with the ion plasma frequency $\omega_i$, which produced the ion Debye length $\lambda_{Di}$. Putting $\epsilon_\mathit{ia}=0$ yields the ion acoustic dispersion relation 
 \begin{eqnarray}
\omega_\mathit{ia}^2(\mathbf{k})=\frac{\omega_i^2}{1+1/k^2\lambda_\mathit{De}^2}\bigg[1&+&\frac{3T_i}{T_e}\bigg(1+ k^2\lambda_\mathit{De}^2\bigg) +\cr
&& \\ 
&+&\frac{k^2\lambda_\mathit{De}^2}{1+k^2\lambda_\mathit{De}^2}\frac{\delta N}{N}\bigg] \nonumber
\end{eqnarray}
The last term in the bracket on the right results from a possible nonlinear density modulation $\delta N$. It vanishes in the long wavelength regime $k^2\lambda_\mathit{De}^2\ll 1$. In order to proceed, we need a treatable form of $\epsilon_\mathit{ia}^{-1}$. It is not difficult to show that this can conveniently be written as
\begin{equation}\label{eq-iaeps}
\frac{1}{\epsilon_\mathit{ia}(\omega,\mathbf{k})}= \frac{k^2\lambda_\mathit{De}^2}{1+k^2\lambda_\mathit{De}^2} \bigg( 1+\frac{\omega_{ia}^2(\mathbf{k}) }{ \omega^2-\omega_\mathit{ia}^2(\mathbf{k}) }\bigg)
\end{equation}
which separates it into two parts. The first term is independent of the presence of ion acoustic waves. It is thus completely spherically symmetric resulting in the known Debye-screening potential field of the point charge \citep[treated in][]{neufeld1955}. Its contribution to the potential at distances $r\gg\lambda_\mathit{De}$, large with respect to the Debye radius, is exponentially small. The nonlinear term in the wave dispersion relation is of higher order and can be neglected meaning that the nonlinear modulation of the wave spectrum is of to large scale for causing a first order effect in the potential disturbance. The dominant effect of the test electron interaction with the ion-acoustic wave spectrum is contained in the wave-mediated part. Its main contribution comes from the resonant denominator in frequency space the contribution of which dominates over that of the exponentially decreasing screened repulsing Coulomb potential $\Phi_D$ outside the Debye sphere. 

Eq. (\ref{eq-iaeps}) inserted into the general expression for the test particle potential Eq. (\ref{eq-potential})  yields for the  wave-induced contribution 
\begin{eqnarray}\label{eq-zero}
\Phi_\mathit{ia}(\mathbf{x},t) &=& \frac{e\lambda_\mathit{De}^2}{16\pi^2\epsilon_0}\int\ \frac{\mathrm{d}\mathbf{k}\ \mathrm{d}\omega\ \omega_\mathit{ia}\,(\mathbf{k})\mathrm{e}^{i\mathbf{k\cdot r}}}{1+k^2\lambda_\mathit{De}^2} \times  \cr
&& \\
&&~~~~~~~ \times\ \left[\frac{\delta(\omega-\mathbf{k\cdot v})}{\omega-\omega_\mathit{ia}(\mathbf{k})}-\frac{\delta(\omega-\mathbf{k\cdot v})}{\omega+\omega_\mathit{ia}(\mathbf{k})}\right] \nonumber
\end{eqnarray}
The argument of the $\delta$-function depends on the direction of electron velocity $\mathbf{v}=v\mathbf{z}$ which we arbitrarily chose in $z$ direction. It is then appropriate to treat the integral in cylindrical rather than spherical coordinates with wave number $k_z$ parallel to the electron velocity $\mathbf{v}$ and $k_\perp$ perpendicular to it. With $\rho$ the radius in the plane perpendicular to $\mathbf{v}$, the argument of the exponential becomes $i\mathbf{k\cdot r} = ik_z(z-vt) +ik_\perp\rho\sin\phi$. Referring to the definition of Bessel functions, the integration with respect to the azimuthal angle $\phi$ results in the Bessel function of zero order, and the expression for the potential reads
\begin{eqnarray}\label{eq-first}
\Phi_\mathit{ia}(z,\rho,t)&=& \frac{e\lambda_\mathit{De}^2}{16\pi\epsilon_0}\int\frac{\mathrm{d}k_z\mathrm{d}k_\perp^2J_0(k_\perp\rho)\,\omega_\mathit{ia}(k_z,k_\perp) }{1+k_z^2\lambda_\mathit{De}^2+k_\perp^2\lambda_\mathit{De}^2}\ \times\cr
&& \\
&\times&\ \left[\frac{\delta(\omega-{k_z v})}{\omega-\omega_\mathit{ia}(\mathbf{k})}-\frac{\delta(\omega-{k_z v})}{\omega+\omega_\mathit{ia}(\mathbf{k})}\right]\mathrm{e}^{ik_z(z-vt)}\mathrm{d}\omega \nonumber
\end{eqnarray}
$\Phi_{ia}$ offers a possible change in sign which opens up the possibility for the potential of becoming attractive for another electron, in which case two electrons may form pairs.  

One first makes use of the $\delta$-functions to replace $k_z=\omega/v$ in the exponential and elsewhere by performing the $k_z$ integration. The two singularities at $\omega=\pm\omega_\mathit{ia}$ require performing the $\omega$-integration in the complex $\omega=\Re(\omega)+i\Im(\omega)$ plane via the principal values of the two integrals and the two residua with integration contour now closed in the lower half-plane, i.e. for $z-vt<0$ and damped ion-acoustic waves $\Im(\omega)<0$. One readily shows that the principal value vanishes, for each integral contributes $\lim_{\epsilon\to0}(\ln\epsilon -\ln\epsilon)+i\pi=i\pi$ which cancel when subtracted. The residua yield the resonant result
\begin{eqnarray}\label{eq-sec}
\Phi_\mathit{ia}(z,\rho,t)=\frac{e\lambda_\mathit{De}^2}{8v\epsilon_0}&\int&\frac{\mathrm{d}k_\perp^2J_0(k_\perp\rho)\,\omega_\mathit{ia}(v,k_\perp) }{1+\omega_\mathit{ia}^2\lambda_\mathit{De}^2/v^2+k_\perp^2\lambda_\mathit{De}^2}\ \times\cr 
 &&\\ 
 &\times& \sin\bigg[\omega_\mathit{ia}(v,k_\perp)\ \bigg(\frac{z}{v}-t\bigg)\bigg]\nonumber
\end{eqnarray}
for the wave-particle interaction part of the electrostatic potential. In this expression the ion acoustic frequency is implicitly defined through the ion-acoustic dispersion relation. Replacing $k_z=\omega_{ia}/v$ the latter can be iterated, yielding to lowest order in the long wavelength regime $k_\perp\lambda_\mathit{De}\ll 1$ that $\omega_\mathit{ia}^2(v,k_\perp)\approx \omega_i^2\lambda^2_\mathit{De}k^2_\perp[1+\omega_\mathit{ia}^2/k_\perp^2v^2)]\approx c^2_\mathit{ia}k^2_\perp/[1-(m_e/m_i)T_e/K_\mathit{test}]$. Here $K_\mathit{test}=m_ev^2/2$ is the test particle kinetic energy. This becomes simply $ \omega_\mathit{ia}^2(v,k_\perp)\approx k^2_\perp c^2_\mathit{ia}/(1-c_\mathit{ia}^2/v^2)$ with $c_\mathit{ia}\approx\omega_i\lambda_\mathit{De}$ the ion sound velocity.  

The wave number integral must be truncated at the Debye radius $k_\perp\lambda_\mathit{De}\leq1$ for the reason that inside the Debye sphere the point charge potential dominates. This accounts for long wavelengths only. Then the integral becomes
\begin{equation}
\Phi_\mathit{ia}(z,\rho,t) \approx C \int\limits_0^1 \frac{\mathrm{d}\xi\, \xi^2\,J_0(\xi\bar\rho)}{1+\xi^2[1+1/(v^2/c_\mathit{ia}^2-1)]}\ \sin(\beta\xi)
\end{equation}
with $\xi=k_\perp/\lambda_\mathit{De}$, $\bar\rho=\rho/\lambda_\mathit{De}$, and $\beta=\zeta(c_\mathit{ia}/v)(1-c_\mathit{ia}^2/v^2)^{-1/2}$, $\zeta=(z-vt)/\lambda_\mathit{De}$. The constant is $C=(e/4\epsilon_0\lambda_\mathit{De})(c_\mathit{ia}/v)(1-c_\mathit{ia}^2/v^2)^{-1/2}$.  Strictly speaking, this integral with respect to $\xi$ is the sum of its principal value and the contribution of the poles at $\xi_\pm=\pm i (1-c_\mathit{ia}^2/v^2)^{1/2}$. At sufficiently large particle speeds the pole contribution is negligible, and only the principal value counts. This is seen as follows. For resonant particles $v\gtrsim c_\mathit{ia}$ the poles are purely imaginary. Extending the singular integral over the entire domain implies that only the positive pole contributes, which is obvious already from Eq. (\ref{eq-zero}) since the exponential vanishes at large $\mathbf{r}$ for positive imaginary part of $k_\perp$ only. In performing the path integration the pole is surrounded in negative direction. Taking the residuum yields a term
\begin{eqnarray}
2\pi i |\xi_+|^2J_0\big(\, i|\xi_+|\bar\rho\big)\ \sin\ \big(i\beta|\xi_+|\big)\ &=& \\ 
-2\pi |\xi_+|^2 I_0\big(\,|\xi_+|&\bar\rho&\!\big)\sinh\ \big(\beta|\xi_+|\big) \nonumber
\end{eqnarray}
$I_0(x)$ is the zero-order modified Bessel function. It is obvious that this entire term for particles close to resonance with $v\gtrsim c_\mathit{ia}, |\xi_+|\sim O(v^2-c_\mathit{ia}^2)$ is very small, confirming that it can safely be neglected.

When calculating the principal part of the integral, we consider the case $k_\perp\rho<1$, i.e. radial distances perpendicular to the particle velocity less than the ion-acoustic wavelength but large with respect to the Debye length. Shortest distances are thus $\bar\rho=1$, yielding $J_0(\xi\bar\rho)=J_0(\xi)$ a function of the integration variable only, varying in the interval $0.77<J_0(\xi)<1.0$. Its average is $\langle J_0\rangle\approx 0.85$ which we extract from the integral
\begin{eqnarray}
\Phi_\mathit{ia}(z,1,t)&\approx& C'\int_0^1\xi^2\mathrm{d}\xi\,\sin(\beta\xi)\cr
&=&\frac{C'}{\beta^3}\bigg[\bigg(1-\textstyle{\frac{1}{2}}\beta^2\bigg)\cos\beta+\beta\sin\beta-1\bigg]
\end{eqnarray}
with $C'=C\langle J_0\rangle$. The integral is of the same form as in \citep{nambu1985}. The requirement $\zeta=z-vt<0$ implies $\beta<0$. The potential becomes negative whenever the expression in the brackets is positive. The interesting case is when the test particle moves at velocity $v\gtrsim c_\mathit{ia}$ exceeding the wave velocity only slightly. Then $|\beta|\ \mathrm{mod}\ {2\pi}>1$, and the dominant term is $\beta^2\cos\beta$ thus confirming \citet{nambu1985} and yielding 
\begin{equation}
\Phi_\mathit{ia}(z,1,t)\approx \big(\big|C'/\beta\big|\big)\cos\beta
\end{equation}
The potential is attractive in all regions $\cos\beta<0$, i.e. $\beta>\pi/2$. In the moving particle frame $\Delta\equiv (z-vt)/\lambda_\mathit{De}$ the potential is attractive behind the particle in its wake in regions $\Delta<0$ for $v\gtrsim c_\mathit{ia}$. Here the value of the potential is
\begin{eqnarray}\label{eq-finia}
\Phi_\mathit{ia}(\Delta)\approx &-&\frac{e\langle J_0\rangle}{4\pi\epsilon_0\lambda_\mathit{De}|\Delta|} \bigg| \cos\bigg[\frac{c_\mathit{ia}\Delta}{(v^2-c_\mathit{ia}^2)^{1/2}}\bigg]\bigg|, \\
&& \frac{\pi}{2}\lesssim\frac{c_\mathit{ia}\big|\Delta\big|}{\sqrt{v^2-c_\mathit{ia}^2}}\lesssim\frac{3\pi}{2}
\end{eqnarray}
The effective distance $|z-vt|\sim\lambda_\mathit{De}$ over which the potential is attractive is thus given by $\ell_\mathit{att}/\lambda_\mathit{De}\approx (v^2/c^2_\mathit{ia}-1)^{1/2}>2/\pi,\ \mathrm{mod}\ 2\pi$, i.e. the attraction is strongest just outside the Debye length which implies that two mutually attaching Debye spheres attract each other. For the resonant particle velocity this condition yields $v_\mathit{res}>1.1\,c_\mathit{ia}$.

In order to attract another electron, it is clear that the two electrons must move close to each other within a distance $\Delta>1$ in the region of negative $\Phi_\mathit{ia}$ both being in resonance with the wave at velocities $v\gtrsim c_\mathit{ia}$. In this case they can form pairs effectively becoming Bosons of either zero or integer spin. We may note that in a magnetic field $\mathbf{B}$ with the electrons moving along the field the ion-sound wave depends on the propagation angle $\cos\theta= \mathbf{k\cdot B}/kB$. In this case we have for the sound speed $c_\mathit{ia}\to c_\mathit{ia}\cos\theta$, and the potential becomes a sensitive function of $\theta$, maximizing along $\mathbf{B}$. Moreover, we can set $\rho=0$ and $\langle J_0(0)\rangle=1$ as only the distance $z$ along $\mathbf{B}$ comes into play.

This attractive potential has to be compared to the wave potential $\Phi_w$ the particles are in resonance with. With the assumption $k\lambda_\mathit{De}\ll 1$ we are in the long-wavelength regime with the potential assumed being nearly constant over the range of variability of the attractive potential. Thus the attractive force of the trapping wave potential is small. In negative wave phases it adds to that of the particle by confining low energy electrons in the potential well. These electrons oscillate at the high trapping frequency with their average speeds in resonance with the wave. Wave trapping though being different in the average helps attracting as in the attracting potential only the average trapped speed $\langle v\rangle\approx c_\mathit{ia}$ counts. The high jitter speed at trapping frequency of the electrons averages out. 

Wave-trapped electrons are the best candidates for forming pairs. Moreover, since a pair of charge $2e$ that has been formed in the negative wave potential well may by the same mechanism produce a negative pair potential $\Phi_\mathit{pair}=\frac{2}{3}\Phi_\mathit{ia}$ over the distance of $3\lambda_\mathit{De}$, it may attract other electrons or pairs to form larger macroparticles of large mass and charge but constant mass-to-charge ratio. In the extreme (though possibly unrealistic) case the maximum number of coagulated electrons could bout equal the number of electrons trapped in the wave potential well, since all the negative potentials of the particles involved into producing attractive potentials add to the wave potential. In effect this mechanism may produce macro-electrons of large mass and charge which behave like a single particle and have such properties as exploited in small mass-ratio numerical PIC simulations.

\subsection{Electron-acoustic pairing potential 
}
Another wave of similar dispersion is the electron-acoustic wave. It is excited wherever the plasma contains two electron populations of different temperatures and densities. It response function resembles that of ion acoustic waves with the only difference that two populations of electrons are involved, and ions are assumed forming a fixed charge neutralizing background such that for the densities$N_i=N_c+N_h$ where indices $c,h$ refer to the cold and hot electron components. Electron acoustic waves are high-frequency waves in the sense that $kv_h,kv_c\ll |\omega-\mathbf{k\cdot u)_c}|$, with $v_c,v_h$ the thermal speeds of the different electron components. The electron-acoustic dielectric response function in its simplest form reads
\begin{equation}
\epsilon_\mathit{ea}(\mathbf{k},\omega)= 1+\frac{1}{k^2\lambda_\mathit{Dh}^2}-\frac{\omega_c^2}{(\omega-\mathbf{k\cdot u}_c)^2}
\end{equation}
The Debye radius for sufficiently large temperature differences $T_h>T_c$ is completely determined by the hot component, and for fixed ions there is no need to include the in term. $\mathbf{u}_c$ is the bulk streaming velocity of cold electrons. The inverse of the dielectric function can again been brought into the same form as for ion-acoustic waves
\begin{equation}
\frac{1}{\epsilon_\mathit{ea}(\mathbf{k},\omega)}\approx\frac{k^2\lambda_h^2}{1+k^2\lambda_h^2}\bigg(1+\frac{\omega_\mathit{ea}^2}{(\omega-\mathbf{k\cdot u}_c)^2-\omega_\mathit{ea}^2}\bigg)
\end{equation}
This is exactly the same form as for ion acoustic waves, however, now with the electron acoustic dispersion relation $\omega_\mathit{ea}^2=k^2c_\mathit{ea}^2/(1+k^2\lambda_h^2)$ and $c_\mathit{ea}^2=v_h^2(N_c/N_h)$. For this reason, the analysis is the same as for the ion-acoustic wave. The result has already been given by \citet{shukla1997} and is listed here for completeness only:
\begin{equation}
\Phi_\mathit{ea}\propto \big(e/|z-ut|\big)\cos\big[|z-vt|/\lambda_h(1-c_\mathit{ea}^2/v^2)^{1/2}\big]
\end{equation}
The bulk speed of the electrons has been suppressed here. As before, there are some ranges in which the wave potential at the test charge can be negative and thus attract other electrons. This will, however, only happen in a plasma where two widely separated in temperature electron populations exist of which the colder one is streaming. interestingly this might be the case in conditions when BGK electron hole modes are excited. In this case the hole generates a dilute hot electron component which is traversed by a rather cold component of beam electrons. Possibly in this case mutually attracting electrons become possible. Unfortunately, electron acoustic waves have not been detected in these cases, however, in numerical simulations of electron hole formation. As electron acoustic waves require strong forcing in order to overcome damping, electron acoustic waves are not a primary candidate for generating attractive wave potentials. 

\subsection{Lower-hybrid pairing potential}
A most important medium frequency wave is the lower-hybrid mode \citep{huba1977,yoon2002}. It propagates in a plasma under almost all conditions on scales below the ion cyclotron radius and frequency. Hence the ions behave non-magnetically while the electrons are completely magnetized being tied to the magnetic field and drifting in the electric field of the wave mode. Lower hybrid waves can be excited by density gradients, diamagnetic drifts and all kinds of transverse currents  $\mathbf{J}_\perp= \sum_sq_sN_s\mathbf{u_{s\perp}}$ in a plasma, where $\mathbf{u}_{s\perp}$ is the perpendicular drift velocity of species $s$. They are primarily electrostatic, propagating at oblique angle with respect to the magnetic field though being strongly inclined with $k_\|< k_\perp$. Their response function including the test particle Coulomb potential term reads
\begin{eqnarray}
\epsilon_\mathit{lh} (\mathbf{k},\omega)= 1+\frac{1}{k^2\lambda_\mathit{De}^2}&-&\frac{\omega_\mathit{lh}^2}{\omega^2}\bigg[1+\frac{m_i}{m_e}\frac{k_\|^2}{k_\perp^2}\ + \cr
&+&~\frac{3k^2}{2k_\perp^2}\ \bigg(1+\frac{\omega_e^2}{\omega_\mathit{ce}^2}\bigg)\ k^2\lambda_\mathit{Di}^2\bigg], \\
&&~~~~~~~~~k_\|/k_\perp\ \approx\ \sqrt{m_e/m_i}\nonumber
 \end{eqnarray}
The lower-hybrid frequency is defined as $\omega_\mathit{lh}^2=\omega_i^2(1+\omega_e^2/\omega_\mathit{ce}^2)^{-1}$. The term in brackets results from the large argument expansion of the derivative of the plasma dispersion function $Z'(\zeta_i)=-2[1+\zeta_iZ(\zeta_i)]$ with $\zeta=\omega/(\omega_ik\lambda_\mathit{Di})$ the argument for the immobile ions. This response function is formally of the same structure as the ion-acoustic response function Eq. (\ref{eq-respia}). Thus defining
\begin{eqnarray}\label{eq-lh}
\omega_{lh}^2(\mathbf{k})=\frac{\omega_\mathit{lh}^2}{1+1/k^2\lambda_\mathit{De}^2}\bigg[1&+&\frac{m_i}{m_e}\frac{k_\|^2}{k_\perp^2}+ \\ 
&+&\frac{3k^2}{2k_\perp^2}\bigg(1+\frac{\omega_e^2}{\omega_\mathit{ce}^2}\bigg)\ k^2\lambda_\mathit{Di}^2\bigg] \nonumber
\end{eqnarray}
the whole formalism developed for ion acoustic waves can be applied to lower-hybrid waves. We write for the inverse response function
\begin{equation}
\frac{1}{\epsilon_\mathit{lh}(\omega,\mathbf{k})}=\frac{k^2\lambda_\mathit{De}^2}{1+k^2\lambda_\mathit{De}^2}\bigg(1+\frac{\omega_\mathit{lh}^2(\mathbf{k})}{\omega^2-\omega_\mathit{lh}^2(\mathbf{k})}\bigg)
\end{equation}
Again, the Debye-screening term outside the brackets is of no interest at distances $r>\lambda_\mathit{De}$. The contribution to the wake potential comes from the integral in Eq. (\ref{eq-first}) with $\omega_\mathit{ia}(\mathbf{k})$ replaced by $\omega_\mathit{lh}(\mathbf{k})$ and $k_z\equiv k_\|=\omega/v\sqrt{\mu}$, where $k_\|\sim k_\perp\sqrt{\mu}$, $\mu=m_e/m_i$, for nearly perpendicular wave propagation. Performing the $\omega$-integration reproduces a form similar to Eq. (\ref{eq-sec}) 
\begin{eqnarray}
\Phi_\mathit{lh}(z,\rho,t)&=&\frac{e\lambda_\mathit{De}^2}{8v\epsilon_0}\int\frac{\mathrm{d}k_\perp^2J_0(k_\perp\rho)\omega_\mathit{lh}(v,k_\perp)}{1+\lambda_\mathit{De}^2\omega_\mathit{lh}^2(v,k_\perp)/\mu v^2+k_\perp^2\lambda_\mathit{De}^2}\ \times\cr
&\times&\ \sin\bigg[\frac{\lambda_\mathit{De}\,\omega_\mathit{lh}(v,k_\perp)}{\sqrt{\mu}\,v}\frac{(z-vt)}{\lambda_\mathit{De}}\bigg]
\end{eqnarray}
but now including the more complicated lower-hybrid frequency Eq. (\ref{eq-lh}). We simplify the lower-hybrid frequency by observing $k_\|^2/\mu k_\perp^2\sim1$. In dense plasma the last term in the brackets becomes $k_\perp^2\lambda_\mathit{Di}^2(\omega_e^2/\omega_\mathit{ce}^2)\sim k_\perp^2 r_\mathit{ce}^2$ which is of the order of the electron gyroradius-to-wavelength squared, being small for completely magnetized electrons. Hence, $\omega_\mathit{lh}^2\lambda_\mathit{De}^2\sim 2V_A^2(v_e/c)^2 \equiv  c^2_\mathit{lh}$ will be used in the factor in front of the sin-function. The lower-hybrid wave in this case propagates at the Alfv\'en speed $V_A$ corrected by the ratio of electron thermal to light velocity. In this approximation and with $\xi=k_\perp\lambda_\mathit{De}$ we have for the lower-hybrid dispersion relation
\begin{eqnarray}
\lambda_\mathit{De}^2\omega_\mathit{lh}^2(v,k)&\approx& \frac{c_\mathit{lh}^2\,\xi^2(1+\omega^2\lambda_\mathit{De}^2/v^2\xi^2)}{1+k^2\lambda_\mathit{De}^2}\bigg(1+\frac{k_\|^2}{\mu k_\perp^2}\bigg)\cr
&\approx& \frac{2c_\mathit{lh}^2\,\xi^2}{1-c_\mathit{lh}^2/v^2}
\end{eqnarray}
which is to be used in the above integral in the long-wavelength approximation $k_\perp\lambda_\mathit{De}\equiv \xi<1$ and $v\gtrsim c_\mathit{lh,\|}=c_\mathit{lh}/\sqrt{\mu}>c_\mathit{lh}$. The last version on the right results from iterating the frequency $\omega=\omega_\mathit{lh}(v,k)$. Within these approximations and restricting to the interval $\xi\lesssim 1$ for long wavelengths, the potential becomes
\begin{eqnarray}
\Phi_\mathit{lh}(z,\bar\rho,t)&\approx& C_\mathit{lh}\int\frac{\xi^2\,\mathrm{d}\xi\,J_0(\xi\bar\rho)}{1+\xi^2(1+2c_\mathit{lh,\|}^2/v^2)}\sin\big(\beta_\mathit{lh}\xi\big)\cr 
&\approx& C'_\mathit{lh}\int\limits_0^{1/\sqrt{3}}\xi^2\mathrm{d}{\xi}\sin(\beta_\mathit{lh}\xi)\\ 
C_\mathit{lh}=\frac{C'_\mathit{lh}}{\langle J_0\rangle} &=&\frac{e}{2\epsilon_0\lambda_\mathit{De}}\frac{c_\mathit{lh}}{(v^2-c_\mathit{lh}^2)^{1/2}},\\ 
&& \beta_\mathit{lh}\equiv \frac{2c_\mathit{lh,\|}\Delta}{(v^2-c_\mathit{lh,\|}^2)^{1/2}} <0 \nonumber
\end{eqnarray}
where $\Delta\equiv (z-vt)/\lambda_\mathit{De}$. One may note that in the only interesting long wavelength regime the factor multiplying $\xi^2$ in the denominator is at most 3. In order to neglect the entire term $\xi^2(1+2c_\mathit{lh}^2/v^2)\ll 1$ and being able to analytically solve the integral one thus requires that the upper limit of the integral is taken as $\xi<1/\sqrt{3}$. Averaging the Bessel function over this interval again produces the numerical factor $\langle J_0\rangle$.  In the argument of the sin-function the larger parallel wave velocity $c_\mathit{lh,\|}= c_\mathit{lh}/\sqrt{\mu}$ appears. It is due to the higher phase velocities of the lower-hybrid waves parallel than perpendicular to the magnetic field, while the test electron moves along the magnetic field at velocity $v\gtrsim c_\mathit{lh,\|}$ being in resonance with the wave. 

With these assumptions the integration of the sin-function with respect to $k_\perp$ can be performed as before and an attractive wake potential is obtained under similar conditions as for the ion-acoustic wave Eq. (\ref{eq-finia}):
\begin{eqnarray}
\Phi_\mathit{lh}(\Delta)&\approx& \frac{C'_\mathit{lh}}{3|\beta_\mathit{lh}|}\,\cos\,\frac{\beta_\mathit{lh}}{\sqrt{3}}\cr
&\approx&\frac{e\langle J_0\rangle}{12\epsilon_0\lambda_\mathit{De}}\frac{\sqrt{\mu}}{\big|\Delta\big|}\bigg(\frac{v^2-c_\mathit{lh,\|}^2}{v^2-c_\mathit{lh}^2}\bigg)^\frac{1}{2}\ \times \\
&\times&\ \cos\Bigg[\frac{2c_\mathit{lh,\|}\big|\Delta\big|}{\sqrt{3\big(v^2-c_\mathit{lh,\|}^2}\big)}\Bigg]\nonumber
\end{eqnarray}
This potential becomes negative for $\frac{1}{2}\pi<|\beta_\mathit{lh}|/\sqrt{3}<\frac{3}{2}\pi \ \mathrm{mod}\ 2\pi$, in which case it attracts a neighboring parallel electron. An attractive potential requires $v\gtrsim c_\mathit{lh}/\sqrt{\mu}\approx 43c_\mathit{lh}$ in an electron-proton plasma. As a consequence the  fraction under the square root does not shorten out but becomes small of the order of $o\!\left({1-c^2_\mathit{lh,\|}/v^2}\right)\sim O\left(\sqrt{\mu}\right)$. Under the condition  on the argument of the cos-function the amplitude of the potential is of the order of
\begin{equation}
\frac{C'_\mathit{lh}}{3\left| \beta_\mathit{lh}\right|}\lesssim \frac{e\langle J_0\rangle}{3\sqrt{3}\epsilon_0\lambda_\mathit{De}}\frac{c_\mathit{lh}}{v}
\end{equation}
which is small of the order of the ratio $c_\mathit{lh}/v\sim \sqrt{\mu}$. Nevertheless, lower hybrid waves may attract some resonant electrons in parallel motion along the magnetic field.  In the transverse direction any electrons gyrate and thus are insensitive to  attraction. Any potential generated will just cause a cross-field electron drift weakly contributing to local current fluctuations.

\subsection{Buneman mode potential inversion}
A most important plasma wave is the current driven non-magnetic Buneman mode \citep{buneman1958,buneman1959}. It occurs under conditions of collisionless shocks, in collisionless guide field reconnection \citep{drake2003,cattell2005}, and in auroral physics, in all cases producing highly dynamical localized electron structures of the type of BGK modes which trap electrons and cause violent effects in plasma dynamics \citep{newman2001}. Again accounting for the presence of test electrons, the dielectric response function of the Buneman mode is
\begin{equation}
\epsilon(\omega, k)=1+\frac{1}{k^2\lambda_\mathit{De}^2}-\frac{\omega_i^2}{\omega^2}-\frac{\omega_e^2}{(\omega-ku)^2}
\end{equation}
with $u$ the current drift velocity of the electrons, and $k$ the one-dimensional wavenumber. For the Buneman mode one has $k\approx \omega_e/u$ and $\omega_i\ll\omega\ll\omega_e$. Under these conditions the (nonlinear) version of the Buneman response function becomes
\begin{equation}
\epsilon_B(\omega, \mathbf{k})=1+\frac{1}{k^2\lambda_\mathit{De}^2}+\frac{\mu}{2}\bigg(\frac{\omega_e}{\omega}\bigg)^3\bigg(1+\frac{3}{2}\frac{\delta N}{N}\bigg)
\end{equation}
The Buneman dispersion relation is obtained as
\begin{equation}
\omega_B^3(\mathbf{k}) = - \omega_e^3\ \frac{\mu}{2} \frac{k^2\lambda_\mathit{De}^2}{1+k^2\lambda^2_\mathit{De}}\bigg(1+\frac{3}{2}\frac{\delta N}{N}\bigg)
\end{equation}
where we retained the nonlinear modulation term proportional to the density variation $\delta N$. In equilibrium it becomes $\delta N/N=-(\epsilon_0/4m_ic_\mathit{ia}^2N)\left|\delta E_B\right|^2$ which is proportional to the Buneman electric field intensity causing hole formation. In the following this term will be neglected. We note that the solution $\omega_B(\mathbf{k})=\Re(\omega_B)+i\Im(\omega_B)$ has a non-negligible imaginary part which must be taken into account. 
Inverting the response function yields,
\begin{equation}
\frac{1}{\epsilon_B(\omega,\mathbf{k})}=\frac{k^2\lambda_\mathit{De}^2}{1+k^2\lambda_\mathit{De}^2}\bigg(1+\frac{\omega_B^3(\mathbf{k})}{\omega^3-\omega_B^3(\mathbf{k})}\bigg)
\end{equation}
The structure of this function is more complicated that in the ion-acoustic which is due to the higher power in frequency and its imaginary part. This function is to be used in Eq. (\ref{eq-one}). Again, the first term just reproduces the Debye screening and can thus be dropped. In order to treat the integral of the second term, we again assume that the electron moves in $z$ direction at velocity $v$. Rewriting the integral in cylindrical coordinates and replacing $k_\|=\omega/v$ as required by the delta function, we find
\begin{eqnarray}
\Phi_B(z,\rho,t)&=&\frac{e\lambda_\mathit{De}^2}{16\pi\epsilon_0}\int\frac{\mathrm{d}\omega\mathrm{d}k_\perp^2\,J_0(k_\perp\rho)}{1+\omega^2\lambda_\mathit{De}^2/v^2+k_\perp^2\lambda_\mathit{De}^2}\ \times\cr
&&\times \ \frac{\omega_B^3(k_\perp,v)\ \exp\,[i\omega(z-vt)/v]}{\left[\omega^3-\omega_B^3(k_\perp,v)\right]}
\end{eqnarray}
Treating the $\omega$-integral is complicated by the third power of the frequency. It requires expansion of the last term into a Laurent series. Since we know that $\omega_B$ is a solution of the dispersion relation the denominator can be expanded around $\omega=\omega_B$ yielding in the denominator $3\omega_B^2(\omega-\omega_B)\left[1+(\omega-\omega_B)/\omega_B+\frac{1}{2}(\omega-\omega_B)^2/\omega_B^2\right]$. The bracket can then be further expanded. Ultimately applying the residuum theorem, only the first term survives producing
\begin{eqnarray}
\Phi_B(z,\rho,t)&=&\frac{ie\lambda_\mathit{De}^2}{4\epsilon_0}\int\frac{k_\perp\mathrm{d}k_\perp}{1+\omega_B^2\lambda_\mathit{De}^2/v^2+k_\perp^2\lambda_\mathit{De}^2}\times \cr
&\times& J_0\big(k_\perp\rho\big)\,\omega_B\big(k_\perp,v\big)\exp\big[i\omega_B(z-vt)/v\big]
\end{eqnarray}
and we must, for $\Im(\omega)>0$ require that $z-vt>0$ and integrate over the positive frequency half-space. Indeed, solving the dispersion relation still, for completeness, keeping the nonlinear term  we obtain the usual Buneman frequency and growth rate
\begin{eqnarray}
\Re\,(\omega_B)&\approx& \frac{\omega_e}{(1+1/k^2\lambda_\mathit{De}^2)^{1/3}}\left(\frac{\mu}{16}\right)^\frac{1}{3}\left(1+\frac{1}{2}\frac{\delta N}{N}\right),\\ 
\Im\,(\omega_B)&=&\sqrt{3}\ \Re\,(\omega_B)\nonumber
\end{eqnarray}
Hence, electrons in resonance with the wave lag slightly behind the wave. The integral may be written as a derivative with respect to $\zeta=(z-vt)/v$. Further simplifying the denominator and defining $\bar\omega=\omega_e(\mu/16)^{1/3}(1+\delta N/2N)\approx 0.03\omega_e(1+\delta N/2N)$ the integral becomes
\begin{eqnarray}
\Phi_B(z,\bar\rho,t)\approx \frac{e}{4\epsilon_0}\frac{\partial}{\partial\zeta}\int\limits_0^1&\xi\mathrm{d}\xi& J_0(\xi\bar\rho)\ \times \cr
&\times&\exp\left[-\bar\omega\xi^\frac{2}{3}\left(\sqrt{3}-i\right)\zeta\right]
\end{eqnarray}
Changing variables and solving for the integral and restricting to the dominant term we find that
\begin{eqnarray}
\Phi_B(\zeta,\rho=0)\ &\approx&\ \frac{3}{4}\frac{e}{\epsilon_0\bar\omega\zeta}\ \exp\bigg(-\zeta\bar\omega\sqrt{3}\bigg)\ \times\cr 
&&\\
&\times&\  \bigg[\cos\left(\bar\omega\zeta+\frac{\pi}{6}\right)+i\sin\left(\bar\omega\zeta+\frac{\pi}{6}\right)\bigg] \nonumber
\end{eqnarray}
holding for $\zeta>0$. Only the real part of the potential has physical relevance. We thus find that the potential can indeed become attractive when the cos-function is negative, i.e. in the interval $\frac{1}{3}\pi\lesssim \bar\omega\zeta\lesssim \frac{4}{3}\pi$ and for resonant electrons lagging slightly behind the wave. This last condition can also be written
\begin{equation}
\frac{\pi}{3}\ \lesssim\  0.03\ \frac{v_e}{v}\frac{|z-vt|}{\lambda_\mathit{De}}\ \lesssimÊ\ \frac{4\pi}{3}
\end{equation}
Such electrons are presumably trapped in the wave potential well which confines them to the interior of holes generated by the Buneman mode. For the distance on which the potential is attractive the last expression yields
\begin{equation}
\big|\,(z-vt)\big|_\mathit{att}\gtrsim 10\ \pi\ (v/v_e)\lambda_\mathit{De}
\end{equation}
For the Buneman mode one requires that $u>v_e$. Electron holes arising from Buneman modes extend up to few $\sim1000\lambda_\mathit{De}$ \citep{newman2001}. They are thus well capable of allowing trapped slow electrons of velocity in the narrow interval $v_e<v<u$ to experience attracting inter-electron potentials and form pairs. 

Since the Buneman mode is a strong wave in the sense that it grows very fast, it has a profound effect on the plasma which appears as hole formation with $\delta N\neq0$ reacting on the wave. In a simplified theory this reaction can be most easily described by taking the variation of the Buneman frequency $\delta\omega=\delta\Re{\omega_B}$ with respect to both density and wave number \citep{treumann1997}. The latter is varied with respect to $k_B=\omega_e/u$, yielding
\begin{equation}
\delta\omega\approx \Re(\omega_B)\left[\ {\textstyle\frac{1}{3}}\left(\frac{u}{v_e}\right)^2k^2\lambda_e^2 +{\textstyle\frac{1}{2}}\frac{\delta N}{N}\ \right], \qquad \begin{array}{lcl}
 k &<& k_B,\\ v_e &<& u~~ \end{array}
\end{equation}
It is customarily interpreted as an operator equation acting on the Buneman mode electric field envelope $E(z,t)$. This procedure results in a nonlinear Schr\"odinger equation
\begin{equation}
\bigg[\ i\frac{\partial}{\partial\tau}\ +\ {\textstyle\frac{1}{2}}\nabla^2_{\bar z}\ +\ \eta\ \big|E(\bar z,\tau)\big|^2\ \bigg]\,E(\bar z,\tau)=0
\end{equation}
where $\tau=\Re(\omega_B)t,\ \bar z=\sqrt{6}\,\omega_ez/u$. The coefficient $\eta=\epsilon_0/8m_i c_\mathit{ia}^2N$ of the nonlinear term results from the density response of the plasma to the presence of the finite amplitude Buneman wave.

The stationary solution in the comoving frame of the Buneman wave is, in this approximation, a caviton of amplitude $E(\bar z)= E_m/\cosh(\bar z/L)$ of width $L=1/\left(E_m\bar\eta^{1/2}\right)$ and maximum dip amplitude $E_m$. In this comoving frame $\bar\eta=\epsilon_0/8m_i(c_\mathit{ia}-u)^2N\approx\epsilon_0/8m_iu^2N$ for $u\gg c_\mathit{ia}$. Electrons trapped in the cavitons have velocities $v< (\epsilon_0/m_eN)^{1/2}E_m$. Oscillating back and forth in the caviton, electrons in their backward traveling phase of motion are sensitive to attraction. Hole-passing electrons in either direction, on the other hand, are not in resonance and thus do not experience any attraction. 
 \begin{figure*}[t!]
\centerline{\includegraphics[width=0.7\textwidth,clip=]{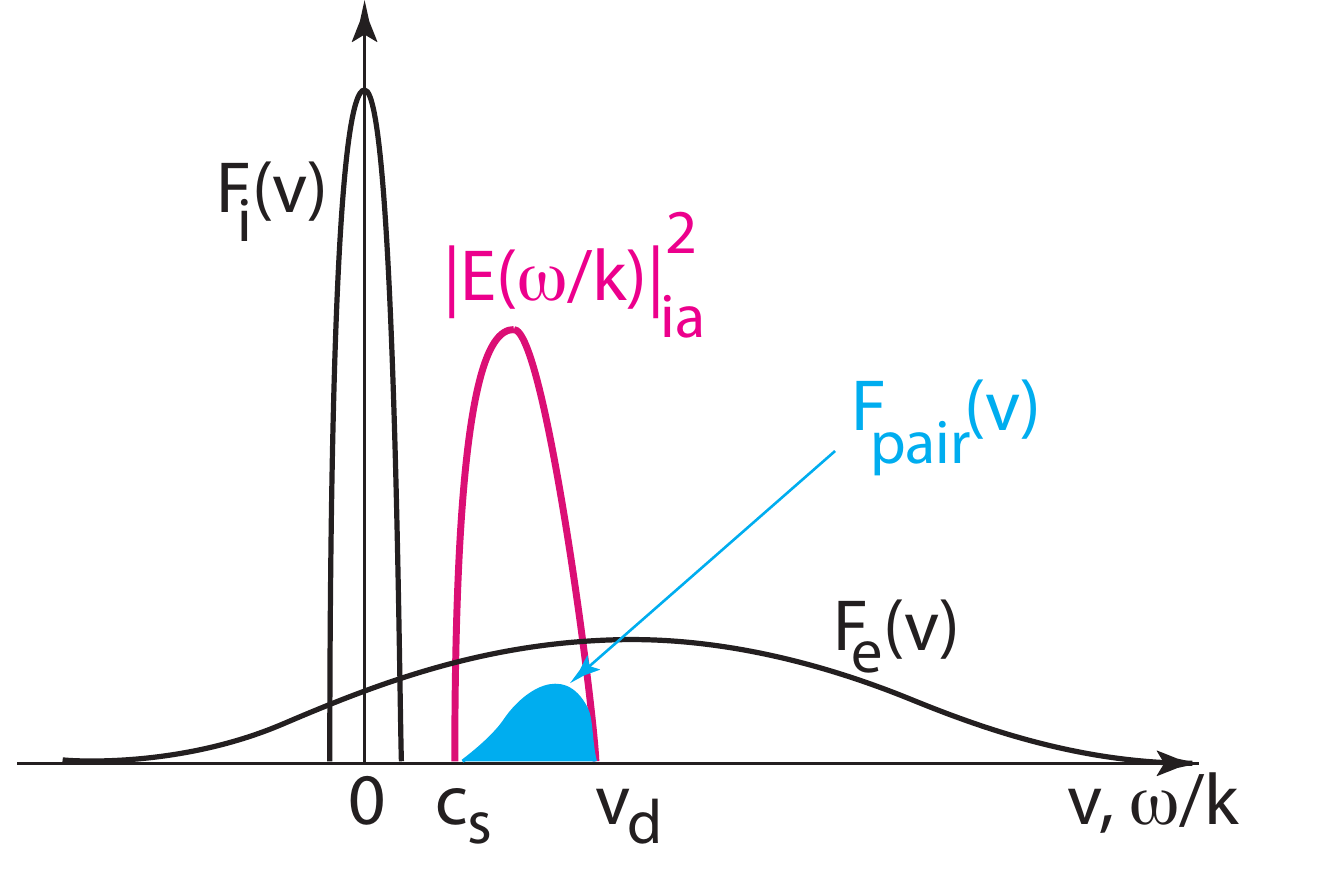} }
\caption[]
{Phase space of ion-acoustic waves excited by the ion-acoustic instability. Shown are the one-dimensional background ion $F_i(v)$ and electron $F_e(v)$ distributions. Ion-acoustic wave with spectrum $|E(\omega/k)|_\mathit{ia}^2$ evolve at phase velocities above the minimum of $c_\mathit{ia}$ in the range $c_\mathit{s}<\omega/k<v_d$. The electron pair distribution function $F_\mathit{pair}(v)$ produced in the high-phase speed range is shown schematically in blue. One may note the very low velocity spread of the pair distribution indicating the much lower pair than original electron background temperature $T_\mathit{pair}\ll T_e$.}\label{fig-pair-a}
\end{figure*}

\section{Conclusions: Possible pairing effects?}
In this paper we examined four types of plasma waves of their capability of causing attraction between two electrons on close distance, finding that all four wave families can under certain conditions contribute to electron pairing. This pairing is a purely classical effect which resembles quantum pairing of electrons in electron-phonon interaction in solid state physics at low temperatures. Nevertheless the mechanisms are similar. This lets one ask whether classical pairing may have observable or unexpected effects. In this last section we present a few speculative hypotheses in this direction.

In solid state physics, electron pair formation is most important at low temperatures.  It is related to super-fluid and super-conducting behavior of matter \citep{huang1987,ketterson1999} in metals and semi-conductors which are based on the fact that pairing electrons become Bosons with either vanishing or integer spin. At low temperatures they behave completely differently from fermionic electrons being capable of releasing their kinetic energy until condensing in their lowest energy level which, in a magnetic field, is the lowest Landau level $\frac{1}{2}\hbar\omega_\mathit{ce}$ \citep{landau1930}. 

In high temperature plasma the particle spin does not play any remarkable role and thermal effects normally inhibit any kind of condensation. However, under completely collisionless conditions with binary collisions excluded, interaction is mediated by plasma waves only. Since pairing is an instantaneous process, pair formation takes place almost immediately if only the necessary and sufficient conditions are satisfied. First of all, electron pairs have mass $m_*=2m_e$. Though the ratio $2e/2m_e=e/m_e$ remains unchanged, electrons become heavy under pairing. The mass increase affects thermal speed, momentum and kinetic energy, doubling both of the latter. 

\begin{figure*}[t!]
\centerline{\includegraphics[width=0.7\textwidth,clip=]{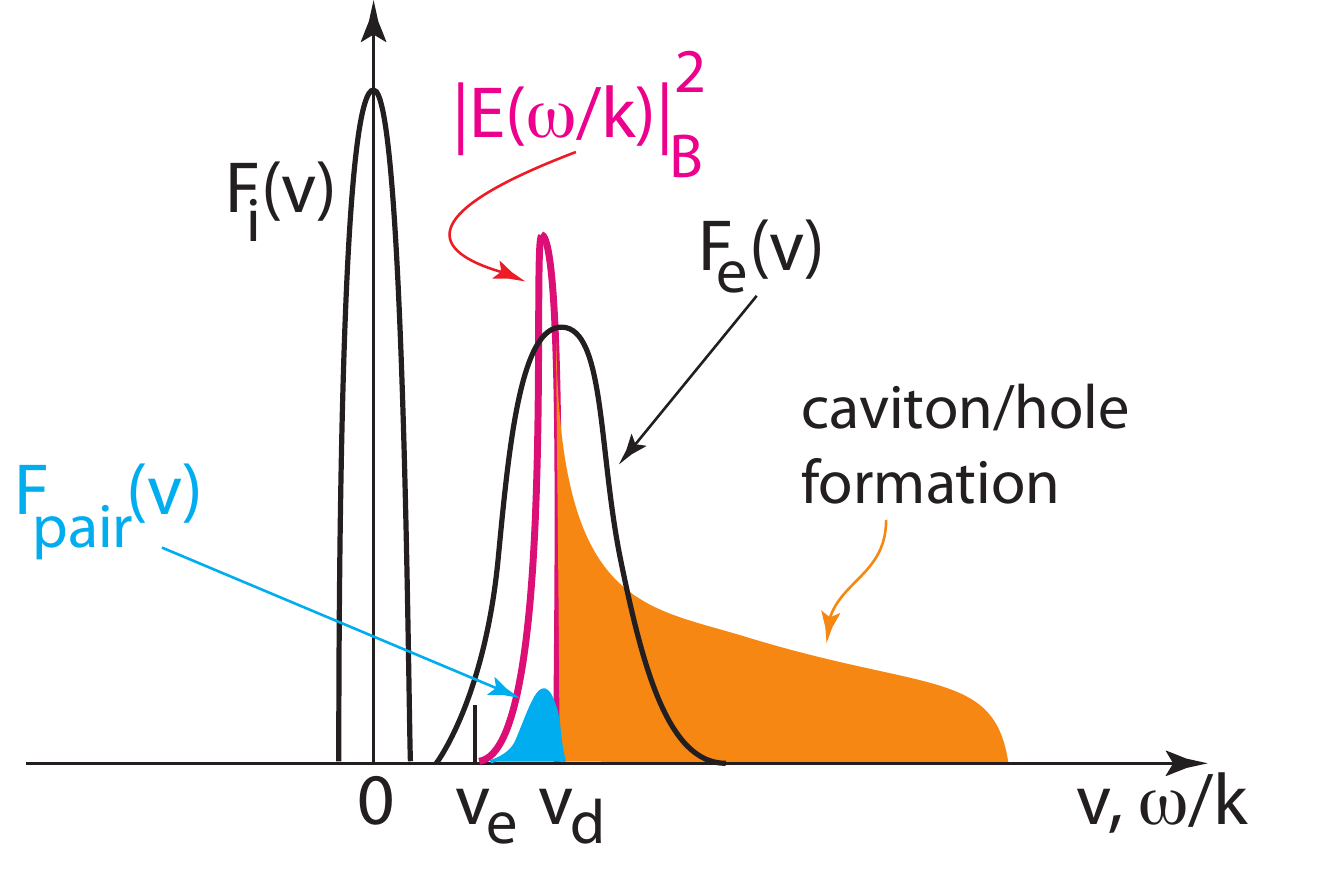} }
\caption[]
{Phase space of Buneman modes excited with spectrum $|E(\omega/k)|_\mathit{b}^2$ and wavenumber $k=\omega_e/v_d$ evolving at phase velocities above $v_e$  for $v_d>v_e$. The spectrum is very narrow in phase velocity. The electron pair distribution function $F_\mathit{pair}(v)$ produced (blue) has similar width as the spectrum and is thus much colder than the original electron distribution. In caviton formation the spectrum extends to much larger phase velocities which, however, has no remarkable effect on the pair distribution. }\label{fig-pair-b}
\end{figure*}

Figure \ref{fig-pair-a} sketches the situation for the case of ion-acoustic waves which may originally have been unstably excited in a thermally imbalanced ion-electron plasma $T_e>T_i$ as shown by the two distributions $F_i(v), F_e(v)$ in one-dimensional phase space. This is the canonical case of ion-acoustic wave excitation. The ion-acoustic wave-spectrum exists in a narrow phase velocity range $c_s<\omega_\mathit{ia}/k<v_d$ as shown in red.  $c_s$ is the minimum of the ion-acoustic wave phase velocity. Attractive potentials can be generated only at substantial wave amplitudes and for electron velocities $v\gtrsim c_\mathit{ia}$. The resulting low density pair distribution is shown in blue. One may note the very narrow velocity spread of the pair distribution $F_\mathit{pair}(v)$ which is at most as wide as the ion-acoustic spectrum. It corresponds to a very low pair temperature $T_\mathit{pair}\ll T_e$ with maximum speed sufficiently far below $v_d$. On may thus conclude, that pair distributions are cold, consisting of heavy electrons and having low speed. Figure \ref{fig-pair-b} is for the Buneman case. The excited spectrum is extremely narrow. Consequently, the pair distribution which is at most as narrow as the spectrum, also has temperature $T_\mathit{pair}\ll T_e$ substantially below the original electron distribution. Caviton formation shown as strong widening of the spectrum in the direction of high phase speeds has no remarkable effect on the pair distribution. It is most interesting, that in all these scenarios the pair plasmas generated turn out to be of very low temperature, much less than the original electron temperature has been. \emph{Pairing in collisionless plasma} thus effectively turns out to be an \emph{efficient non-radiative electron cooling} mechanism.

Since in collisionless plasma electrons escape any collisions and thermal equilibration times are long, the sudden change of particle state from free fermionic electrons to free bosonic pairs might possibly have other unexpected effects. The first is that the plasma after pairing consists of a two-temperature electron plasma of constant charge-to-mass ratio but dilute double-mass cold electrons. Such a two-electron temperature plasma classically excites high frequency/high velocity electron acoustic waves which are radiated away from the pairing region. This is a classical effect independent of the fermionic or bosonic properties. In principle it could be observed if excitation would be strong enough to overcome the strong damping of the electron acoustic waves. 

The second effect may occur in pairing in presence of  ion-acoustic waves. In this case the resonantly produced pairs move at speed slightly \emph{larger} than the phase velocity $c_\mathit{ia}$ of the ion acoustic wave. All pairs created are free particles not obeying any bound states. Once they are created as Bosons, they escape further interaction with the ion-acoustic wave because ion-acoustic waves are phonons which do not interact with Bosons of higher velocity. This physical insight lies at the bottom of Landau's theory of super-fluidity \citep{fetter1971,huang1987}. Pairing electrons in this case behave like a dilute super-fluid flowing on the hot plasma background. In the presence of a slowly variable electric field this bosonic electron fluid becomes superconducting even in the classical limit causing strong current amplification by sucking the current into the superconducting region and giving rise to the occurrence of current filaments. This mechanism requires the presence of a sufficiently strong spectrum of ion-acoustic waves (Figure \ref{fig-pair-a}), which implies electron temperatures $T_e>T_i$, a situation typical for the solar wind and Earth's bow shock (or any collisionless shock in interplanetary space) but untypical for the magnetospheric magnetotail. It might thus be hypothesized that electron acoustic wave-mediated pairing in the solar wind and more generally in collisionless shocks could cause kind of (classical) superconducting domains with unusual amplification of the electric current flow.    

The third hypothetical effect is bosonic condensation which takes place at very low temperatures. Pair temperatures in plasma are indeed low; they are in fact so low that the distributions could be interpreted as a condensate produced classically in high temperature plasma. Whether the pair temperatures become low enough to induce further bosonic condensation, is not known. What concerns this hypothetical condensation so it is very difficult to predict at what energy level it could take place. In zero-temperature solid state matter condensation drops down to the lowest energy state \citep{fetter1971}. In magnetized media this is the mentioned lowest Landau level $\frac{1}{2}\hbar\omega_\mathit{ce}$ \citep{landau1930,huang1987}. In high temperature plasma with $1<T_e< 10^3$ eV this should not be possible, for finite temperature effects make energy levels such low inaccessible. Landau levels in this case form a quasi-continuum. Nevertheless, high Landau levels $T_L\ll T_e$ still far below $T_e$ are in principle accessible for pairs to slide down to and condensate in. So far it cannot be predicted what the lowest accessible Landau level would become. The excess energy released during such a condensation can be given away in excitation of plasma waves. However, if the hypothetical condensation really takes part in an instant then there will be no time available for exciting plasma waves. The excess energy must then be radiated away as electromagnetic radiation. Assume that the plasma had a temperature of $T_e\sim 100$ eV, typical for the solar wind at 1 AU. An energy difference of $T_e-T_\mathit{pair}\approx$ few eV will in this case be emitted most probably in a weak soft-X-ray line of energy about corresponding to the temperature difference.  Weak soft-X ray emission in plasma could in this case result from pairing of a fraction of the electron distribution. In the solar wind such electrons are present as halo electrons. 

In the case of the Buneman instability pairs are formed from the slower electron component (Figure \ref{fig-pair-b}). This electron component is trapped in the cavity/hole that forms in the nonlinear evolution of the Buneman mode. Pairs are cold of temperature far below $T_e$, remain trapped and sink down to the bottom of the cavity potential well which they cannot escape from and where they form a cold massive and non-oscillating trapped bosonic heavy-electron component which is neither super-fluid nor super-conducting but by its negative momentum that causes it to lag behind the Buneman cavity will contribute to steeping of the rear wall of the cavity and cause cavity asymmetry.  No quantum effect is involved in this case which remains completely classical (except for the Fermion-Boson transition) and should therefore be observable. Indeed, numerical simulations \citep{newman2001} indicated asymmetries in Buneman mode generated electron holes. 

It is interesting to speculate on the importance of Buneman induced pairing for reconnection. Guide field simulations and observations indicate that the Buneman mode causes generation of electron holes during reconnection \citep{drake2003,cattell2005}. This is indeed favored in the geomagnetic tail where electron temperatures are lower than ion temperatures inhibiting ion-acoustic wave excitation. Production of a very cold though dilute electron pair plasma in Buneman excited electron holes implies remagnetization of the hole-trapped pairs. During the convective transport of holes into the current sheet the magnetized pair plasma may advect the frozen magnetic field flux tubes into the current sheet, a mechanism which could contribute to enhanced reconnection.




\begin{acknowledgements}
This research was part of a Visiting Scientist Program at ISSI, Bern, executed by RT. Hospitality of the ISSI staff is thankfully acknowledged. 
\end{acknowledgements}











\begin{thebibliography}{99}

\bibitem[Buneman(1958)]{buneman1958} Buneman O (1958) Instability, turbulence, and conductivity in current-carrying plasma, Phys Rev Lett {1}, 8-9, doi:10.1103/PhysRev.Lett.1.8

\bibitem[Buneman(1959)]{buneman1959} Buneman O (1959)  Dissipation of currents in ionized media, Phys Rev {115}, 503-517, doi:10.1103/PhysRev.115.503

\bibitem[Cattell et al.(2005)]{cattell2005} Cattell C, Dombeck J, Wygant J, Drake J~F, Swisdak M, Goldstein M L, Keith W, Fazakerley A, Andr\'e M, Lucek E \& Balogh A  (2005) Cluster observations of electron holes in association with magnetotail reconnection and comparison to simulations, J Geophys Res 110, A01211, doi:10.1029/2004JA010519

\bibitem[Drake et al.(2003)]{drake2003} Drake J~F, Swisdak M, Cattell C, Shay M A, Rogers B N  \& Zeiler A (2003) Formation of electron holes and particle energization during magnetic reconnection, Science 299, 873-877, doi:10.1126/science.1080333

\bibitem[Fetter \& Walecka(1971)]{fetter1971} Fetter A~L \& Walecka J~D  (1971) Quantum theory of many-particle systems (McGraw-Hill Publ. Comp., New York, USA) 

\bibitem[Huang(1987)]{huang1987} Huang K  (1987) Statistical Mechanics, 2nd ed (John Wiley \& Sons, New York, USA) 

\bibitem[Huba et al(1977)]{huba1977} Huba J~D, Gladd N~T \& Papadopoulos K (1977) The lower-hybrid-drift instability as a source of anomalous resistivity for magnetic field line reconnection, Geophys Res Lett 4, 125-128, doi: 10.1029/GL004i003p00125

\bibitem[Ketterson \& Song(1999)]{ketterson1999} Ketterson~J~B \& Song S~N (1999) Superconductivity (Cambridge University Press, Cambridge UK) 

\bibitem[Krall \& Trivelpiece(1973)]{krall1973} Krall N~A \& Trivelpiece A W (1973) Principles of Plasma Physics (McGraw-Hill, New York, USA) 

\bibitem[Landau(1930)]{landau1930} Landau L~D  (1930) Diamagnetismus der Metalle, Zeitschrift f\"ur Physik 64, 629-637

\bibitem[Nambu \& Akama(1985)]{nambu1985} Nambu~M \& Akama~H (1985) Attractive potential between resonant electrons, Phys Fluids 28, 2300-2301, doi:10.1063/1.865284

\bibitem[Nambu \& Nitta(2001)]{nambu2001} Nambu~M \& Nitta~H (2001) On the Shukla-Nambu-Salimullah potential in magnetized plasma, Phys Lett A 300, 82-85, doi:10.1016/S0375-9601(02)00741-7

\bibitem[Neufeld \& Ritchie(1955)]{neufeld1955} Neufeld~J \& Ritchie~R~H (1955) Passage of charged particles through plasma, Phys Rev 98, 1632-1642, doi:10.1103/PhysRev.98.1632

\bibitem[Newman et al.(2001)]{newman2001} Newman D~L, Goldman M~V, Ergun R~E \& Mangeney A  (2001) Formation of Double Layers and Electron Holes in a Current-Driven Space Plasma, Phys Rev Lett 87, 255001, doi:10.1103/PhysRev.Lett.87.255001 

\bibitem[Shukla \& Melands\o(1997)]{shukla1997} Shukla~P~K \& Melands\o~F (1997) Test charge potential in the presence of electron-acoustic waves in plasmas, Phys Scripta 55,239-240, doi:, Phys Lett A 291, 413-416, doi: 10.1088/0031-8949/55/2/013


\bibitem[Shukla et al(2001)]{shukla2001} Shukla~P~K, Nambu~M \& Salimullah~M (2001) Effect of ion polarization drift on shielding and dynamical potentials in magnetized plasmas, Phys Lett A 291, 413-416, doi: 10.1016/S0375-9601(01)00762-9

\bibitem[Treumann \& Baumjohann(1997)]{treumann1997} Treumann~ R~ A \& Baumjohann~W (1997) Advanced Space Plasma Physics (Imperial College Press, London) ch. 11.4

\bibitem[Yoon et al(2002)]{yoon2002} Yoon P~H, Lui A~T~Y \& Sitnov M~I (2002) Generalized lower-hybrid drift instabilities in current-sheet equilibrium, Phys Plasmas 9, 1526-1538, doi:10.1063/1.1466822






\end{thebibliography}
\end{document}